%% file: flares-balmer.tex
\documentclass[fleqn,usenatbib]{mnras}

\usepackage{newtxtext,newtxmath}

\usepackage[T1]{fontenc}
\usepackage{ae,aecompl}

\usepackage{graphicx}	
\usepackage{amsmath}	
\usepackage{amssymb}	
\usepackage[usenames,dvipsnames]{xcolor} 
\usepackage[normalem]{ulem} 
\usepackage{hyperref}

\usepackage{lipsum} 
\usepackage{multirow}

\newcommand{\eagle}{\mbox{\sc{Eagle}}}

\newcommand{\flares}{\mbox{\sc Flares}}

\newcommand{\synthesizer}{\mbox{\tt Synthesizer}}
\newcommand{\spitzer}{\mbox{\it Spitzer}}

\newcommand{\hubble}{\mbox{\it Hubble}}
\newcommand{\jwst}{\mbox{\it JWST}}

\title[FLARES XIV: The Balmer/4000~\AA\ Break]{First Light And Reionisation Epoch Simulations (FLARES) XIV: The Balmer/4000~\AA\ Breaks of Distant Galaxies}

\author[Stephen M. Wilkins et al.]{Stephen M. Wilkins$^{1,2}$\thanks{E-mail: s.wilkins@sussex.ac.uk}, 
Christopher C. Lovell$^{3,1}$, 
Dimitrios Irodotou$^{4}$, 
Aswin P. Vijayan$^{5,6}$, 
\newauthor
Anton Vikaeus$^{7}$, 
Erik Zackrisson$^{7}$, 
Joseph Caruana$^{8,2}$,
Elizabeth R. Stanway$^{9}$,
Christopher J.~Conselice$^{10}$
\newauthor
Louise T. C. Seeyave$^{1}$, 
William J. Roper$^{1}$, 
Katherine Chworowsky$^{11}$,
Steven L. Finkelstein$^{11}$
\\
$^{1}$Astronomy Centre, University of Sussex, Falmer, Brighton BN1 9QH, UK\\
$^{2}$Institute of Space Sciences and Astronomy, University of Malta, Msida MSD 2080, Malta \\
$^{3}$Institute of Cosmology and Gravitation, University of Portsmouth, Burnaby Road, Portsmouth, PO1 3FX, UK\\
$^{4}$Department of Physics, University of Helsinki, Gustaf Hällströmin katu 2, FI-00014, Helsinki, Finland\\
$^{5}$Cosmic Dawn Center (DAWN) \\
$^{6}$DTU-Space, Technical University of Denmark, Elektrovej 327, DK-2800 Kgs. Lyngby, Denmark \\
$^{7}$Observational Astrophysics, Department of Physics and Astronomy, Uppsala University, Box 516, SE-751 20 Uppsala, Sweden \\
$^{8}$Department of Physics, Faculty of Science, University of Malta, Msida MSD 2080, Malta\\
$^{9}$Department of Physics, University of Warwick, Gibbet Hill Road, Coventry, CV4 7AL, UK\\
$^{10}$ Jodrell Bank Centre for Astrophysics, Department of Physics and Astronomy, University of Manchester, Oxford Road, Manchester M13 9PL, UK\\
$^{11}$ Department of Astronomy, The University of Texas at Austin, Austin, TX, USA\\
}
\date{Accepted XXX. Received YYY; in original form ZZZ}

\pubyear{2023}

\begin{document}
\label{firstpage}
\pagerange{\pageref{firstpage}--\pageref{lastpage}}
\maketitle

\begin{abstract}

With the successful launch and commissioning of \jwst\ we are now able to routinely spectroscopically probe the rest-frame optical emission of galaxies at $z>6$ for the first time. Amongst the most useful spectral diagnostics used in the optical is the Balmer/4000~\AA\ break; this is, in principle, a diagnostic of the mean ages of composite stellar populations. However, the Balmer break is also sensitive to the shape of the star formation history, the stellar (and gas) metallicity, the presence of nebular continuum emission, and dust attenuation. In this work we explore the origin of the Balmer/4000~\AA\ break using the \synthesizer\ synthetic observations package. We then make predictions of the Balmer/4000~\AA\ break using the First Light and Reionisation Epoch Simulations (\flares) at $5<z<10$. We find that the average break strength weakly correlates with stellar mass and rest-frame far-UV luminosity, but that this is predominantly driven by dust attenuation. We also find that break strength provides a weak diagnostic of the age but performs better as a means to constrain star formation and stellar mass, alongside the UV and optical luminosity, respectively. 

\end{abstract}




\input Sections/intro

\input Sections/theory

\input Sections/flares

\input Sections/predictions
\input Sections/conclusion

\section*{Acknowledgements}

We would like to thank Adam Carnall and Tobias Looser for providing measurements of the break using our chosen diagnostic.

We thank the \eagle\, team for their efforts in developing the \eagle\, simulation code. 

This work used the DiRAC@Durham facility managed by the Institute for Computational Cosmology on behalf of the STFC DiRAC HPC Facility (www.dirac.ac.uk). The equipment was funded by BEIS capital funding via STFC capital grants ST/K00042X/1, ST/P002293/1, ST/R002371/1 and ST/S002502/1, Durham University and STFC operations grant ST/R000832/1. DiRAC is part of the National e-Infrastructure. 

SMW and WJR thank STFC for support through ST/X001040/1. CCL acknowledges support from a Dennis Sciama fellowship funded by the University of Portsmouth for the Institute of Cosmology and Gravitation. DI acknowledges support by the European Research Council via ERC Consolidator Grant KETJU (no. 818930) and the CSC – IT Center for Science, Finland. APV acknowledges support from the Carlsberg Foundation (grant no CF20-0534). The Cosmic Dawn Center (DAWN) is funded by the Danish National Research Foundation under grant No. 140. AV and EZ acknowledge funding from the Swedish National Space Agency. EZ also acknowledges grant 2022-03804 from the Swedish Research Council. LTCS is supported by an STFC studentship.  

We also wish to acknowledge the following open source software packages used in the analysis: \textsc{Scipy} \cite[][]{2020SciPy-NMeth}, \textsc{Astropy} \cite[][]{robitaille_astropy:_2013}, \textsc{Matplotlib} \cite[][]{Hunter:2007} and WebPlotDigitizer \cite[][]{Rohatgi2020}. 

We list here the roles and contributions of the authors according to the Contributor Roles Taxonomy (CRediT)\footnote{\url{https://credit.niso.org/}}.
\textbf{Stephen M. Wilkins}: Conceptualization, Data curation, Methodology, Investigation, Formal Analysis, Visualization, Writing - original draft.
\textbf{Christopher C. Lovell,  Aswin P. Vijayan}: Data curation, Methodology, Writing - review \& editing.
\textbf{Joseph Caruana, Katherine Chworowsky, Chris Conselice, Steven Finkelstein, Dimitrios Irodotou,  William Roper, Louise T. C. Seeyave, Elizabeth Stanway, Anton Vikaeus, Erik Zackrisson}: Writing - review \& editing.

\section*{Data Availability}

The data associated with the paper will be made publicly available at \href{https://flaresimulations.github.io}{https://flaresimulations.github.io} on the acceptance of the manuscript.



\bibliographystyle{mnras}
\bibliography{flares, flares-balmer, flares-balmer-jwst} 




\appendix

\bsp	
\label{lastpage}
\end{document}

%% file: Sections/intro.tex
\section{Introduction}\label{sec:intro}

A key goal in extragalactic astrophysics is to accurately measure various fundamental physical properties - including the star formation rate, stellar mass, metallicity, and dust attenuation - for the distant, high-redshift ($z>4$), galaxy population. Doing so provides the essential constraints to galaxy formation models, ultimately providing insights into physical processes responsible for early galaxy formation and evolution.

Obtaining robust constraints on these properties requires deep near-infrared observations probing the rest-frame UV and optical emission. Early insights have come from the combination of \hubble\ and \spitzer\ \citep[e.g.][]{Hashimoto2018, DeBarros2019, Endsley2021, Tacchella22}. However, these insights were limited by the low sensitivity and resolution of \spitzer, and contamination from strong line emission \citep[e.g.][]{Stark13, Wilkins2013d, Wilkins2016b, Wilkins2020}. Through its sensitivity, wavelength-coverage, and spectroscopic capabilities, \jwst\ is now overcoming these limitations, with the first spectroscopic measurements now emerging \citep[e.g][]{Schaerer22, Trussler22, Matthee22, Trump23, ArrabalHaro23, Fujimoto23}. Crucially, for this present work, these spectroscopic studies now include detections of the rest-frame UV and optical continuum at sufficient signal-to-noise to enable the measurement of key continuum features \citep[e.g][]{Carnall2023, Bunker23, Cameron23, Looser23, Hsiao2023}.

While nebular emission lines, which provide insights into the inter-stellar medium and sources of ionising photons, are a key target for spectroscopic observations, continuum features also contain invaluable information about the stellar populations, in particular the older aspects of the stellar population. Arguably, the most prized continuum feature is the combined Balmer/4000~\AA\ break feature \citep{Bruzual1983, Poggianti97}. For a \emph{simple} (single-aged) stellar population this feature is strongly sensitive to the age, making it a potentially powerful diagnostic of the star formation histories of galaxies \citep{Worthey94, Poggianti97}.

Previous indications of notably strong Balmer breaks in high redshift galaxies such as \citet{Hashimoto2018,2021Laporte,2020Roberts-Borsani}, prompts further investigation into the underlying mechanisms responsible for producing such strong breaks. Importantly, we seek to discern whether current state of the art hydrodynamic simulations can accommodate such detections -- thus shedding new light on the debate of the onset of star formation in the early Universe.

In this work we begin by exploring the physical properties influencing the strength of this feature. Specifically, we use the \synthesizer\footnote{\url{https://github.com/flaresimulations/synthesizer}} (Lovell et al., \emph{in-prep}) synthetic observations code to explore how the choice of star formation history, metallicity, nebular emission, dust, initial mass function, and choice of stellar population synthesis models affect this observational feature. These are modelled using relatively simple parameterisations, in order to clarify their individual effects on this spectral break. Such features can be obscured in the more complex history and combination of physical properties described by full computational simulations for any given galaxy.

We then make detailed predictions of the Balmer break strength using the First Light And Reionisation Epoch Simulations \citep[\flares,][]{FLARES-I, FLARES-II} for galaxies at high-redshift ($5\le\,z\,\le10$). \flares\ is a suite of hydrodynamical zoom simulations that, by virtue of its strategy, simulates a wide range of galaxies with ($M_{\star}=10^8-10^{11}\ {\rm M_{\odot}}$) at high-redshift ($z>5$). Unlike our simple modelling, which assumes a smooth parameterisation of the star formation history, \flares\ galaxies have realistic, complex star formation and metal enrichment histories and star-dust geometries. \emph{So far}, \flares\ has been shown to produce an excellent match to observational constraints \citep{FLARES-I, FLARES-II, Tacchella22, FLARES-IV, FLARES-VI, FLARES-VII, FLARES-XI}, including early constraints from \jwst\ \citep{FLARES-VII, FLARES-XI, Adams23b}. Using \flares\ we explore how the predicted break strength correlates with stellar mass, far-UV luminosity and key physical properties, including the specific star formation rate, age, dust attenuation, and optical mass-to-light ratio, in galaxies with realistically complex and stochastic star formation histories. 

This paper is structured as follows: we begin, in Section \ref{sec:theory}, to explore the impact of physical properties on the Balmer break using simple star formation and metal enrichment histories. Next, in Section \ref{sec:flares}, we introduce the \flares\ project, including a detailed description of our spectral energy distribution modelling procedure. In Section \ref{sec:predictions} we make predictions for the break strength, including a comparison with current spectroscopic observations (\S\ref{sec:predictions:observations}), and an exploration of how the predicted break strength correlates with and is impacted by various physical properties (\S\ref{sec:predictions:physical}).  We then summarise our findings in this work and present our conclusion in section \S\ref{sec:conc}.

%% file: Sections/theory.tex
\section{Theoretical Background}\label{sec:theory}

In this work we are interested in exploring predictions for the strength of the Balmer/4000~\AA\, break feature. The Balmer break is a sharp feature occurring at the Balmer limit (3645~\AA) and is caused by bound-free absorption by electrons in the electron band $n=2$. This feature is most prevalent in A stars ($1.4-2.1\ {\rm M_{\odot}}$) where effective temperatures of $\approx 10,000$ K maximise the number of electrons in the $n=2$ band. In hotter (OB) stars an increasing fraction of Hydrogen is ionised reducing the absolute number of electrons in $n=2$, while in cooler stars the falling temperature reduces the number in $n=2$. However, the spectra of cooler stars are increasingly affected by line-blanketing. The combination of these features gives rise to a strong dependency slope around 4000~\AA\ to the age of single-aged stellar population. 

A popular metric of the break is $D_{4000}$ defined by \citet{Bruzual1983} using windows at [3750, 3950]\AA\ and [4050, 4250]\AA. However, the blue window of $D_{4000}$ is red-ward of the Balmer limit (3645~\AA) and thus not as useful for quantifying the break in younger galaxies. For this reason we adopt the definition  of \citet{Binggeli2019} and define the break strength using two windows at [3400,3600]\AA\ and [4150,4250]\AA\ instead. These windows are chosen to straddle the Balmer limit, and to also avoid strong nebular line emission while minimising their separation to reduce the impact of dust attenuation. Throughout this work we express the break strength simply as the logarithmic ratio of the red and blue window luminosities.  

For young, star-forming galaxies, dominated by emission from O and B stars, and with little dust attenuation, this quantity is close to zero. For older stellar populations this quantity becomes progressively more positive, with a 1 Gyr old simple stellar population reaching a strength of $\approx 0.5$.

\begin{figure}
	\includegraphics[width=\columnwidth]{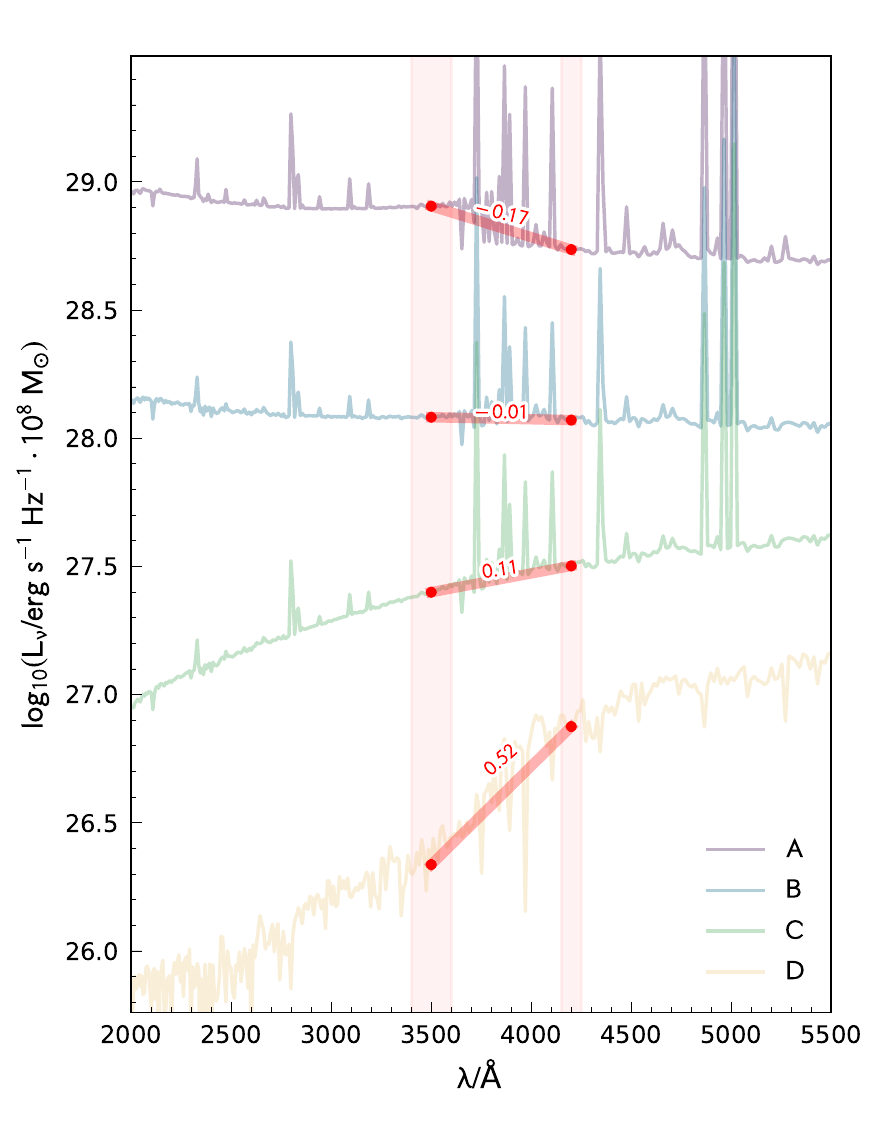}
	\caption{Visualisation of the Balmer break in the context of four simple scenarios, each with the same stellar mass, $M_{\star}=10^{8}\ {\rm M_{\odot}}$. A: 10 Myr constant star formation and no dust. B: 100 Myr constant star formation and no dust. C: 100 Myr constant star formation and V-band optical depth $\tau_{V}=1$. D: 1 Gyr old burst and no dust. Each scenario assumes $Z=0.001$ and no escape of Lyman continuum photons (i.e. $f_{\rm esc}=0$). The two Balmer break windows are denoted by the red vertical bands. \label{fig:theory_sed}}
\end{figure}

To gain insight into the physical properties driving the break strength we utilise a simple toy model. This allows us to explore the sensitivity of the break to the star formation and metal enrichment history, reprocessing by dust and gas, as well as the choice of stellar population synthesis (SPS) model and initial mass function (IMF). For this toy model we employ smooth parametric star formation histories, a single metallicity, and a simple screen model for reprocessing by dust and gas (i.e. stellar populations are all equally affected by dust). By default we assume the same SPS model and IMF as utilised by \flares\ \citep{FLARES-II}: v2.2.1 of the Binary Population And Spectral Synthesis \citep[BPASS][]{BPASS2.2.1} and a \citet{chabrier_galactic_2003} initial mass function extending to $300\ {\rm M_{\odot}}$. To model the contribution of nebular emission we use v17.03 of the \texttt{cloudy} photo-ionisation code \citep[][]{Cloudy17.02}, assuming identical nebular and stellar metallicities, and assume a reference ionisation parameter\footnote{Here the reference ionisation parameter is the ionisation parameter assumed for stellar populations with an age of 1 Myr and $Z=0.01$. For stellar populations with different ages and metallicities the actual assumed ionisation parameter is scalled $\propto Q^{1/3}$ - see \citet{Wilkins2020} and \citet{FLARES-XI} for more details.} of $U_{\rm ref}=0.01$. In Figure  \ref{fig:theory_sed} we use this model to visualise the break for four scenarios: a young 10 Myr burst of dust-free star formation (A), 100 Myr constant star formation with dust (C) and without dust (B), and a 1 Gyr old burst (D). The differences between these models is discussed is the proceeding sections.

\subsection{Star formation history}\label{sec:theory:sfh}

\begin{figure}
	\includegraphics[width=\columnwidth]{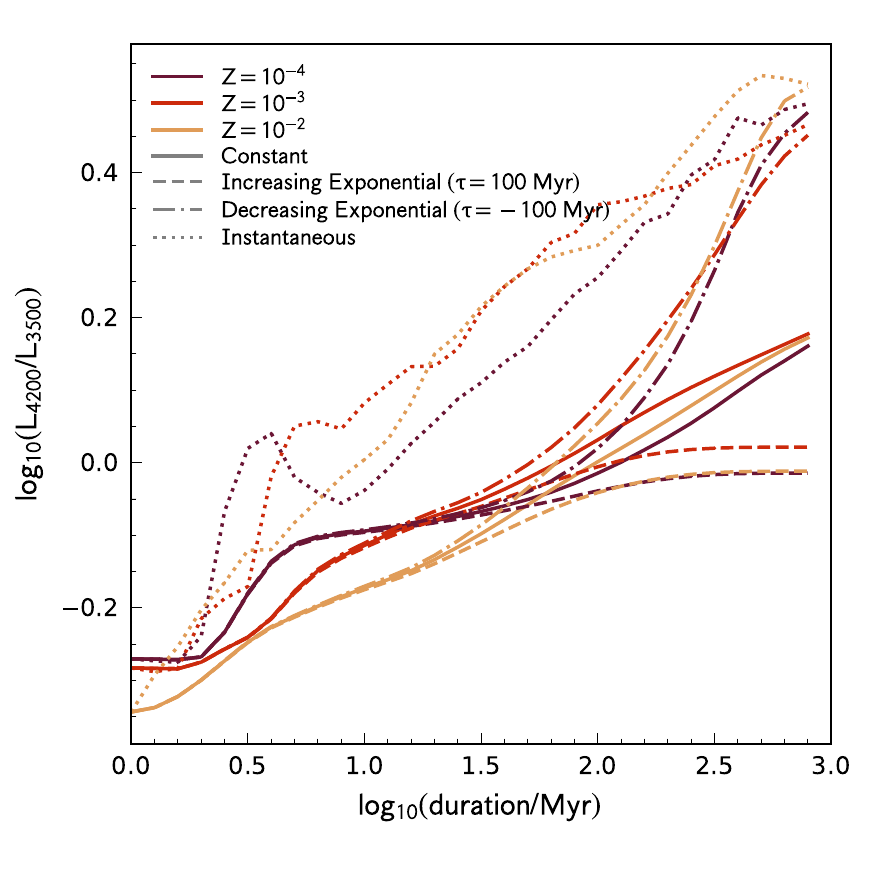}
	\includegraphics[width=\columnwidth]{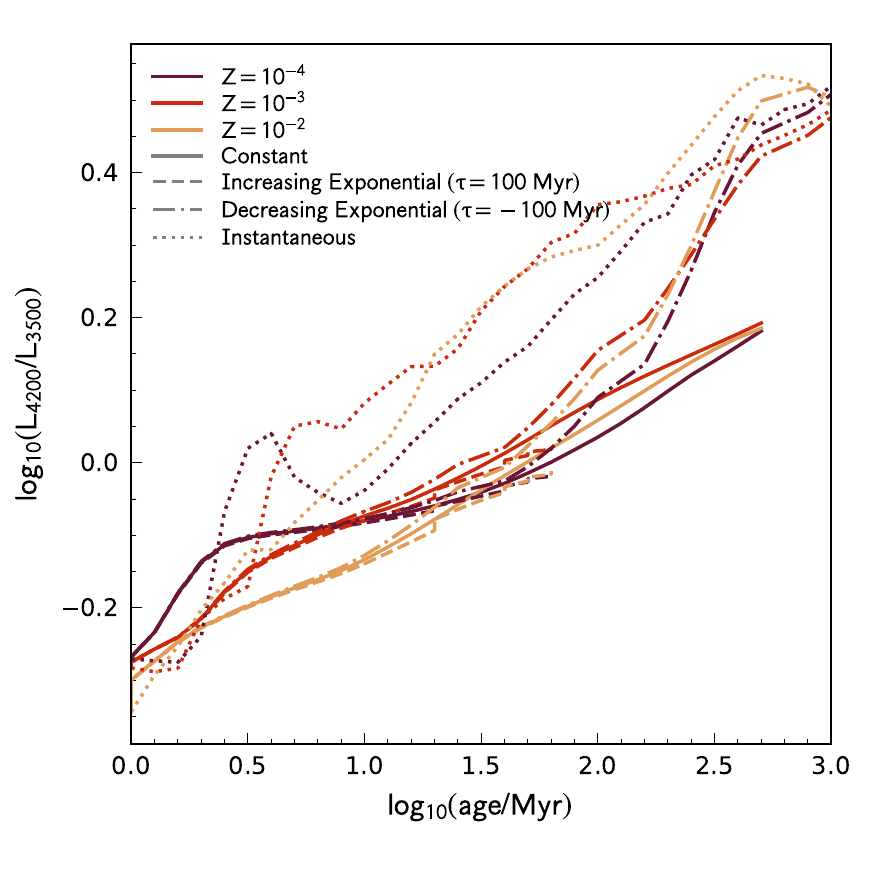}
	\caption{The sensitivity of the break strength to the star formation history and metallicity. In the top panel the $x$-axis is the duration of star formation while in the bottom we use median age instead - for an instantaneous burst the duration and age are equivalent while for a constant star formation history the age is half the duration. In both panels we show results assuming 4 different star formation histories: an instantaneous burst (dotted line), exponentially decreasing star formation (dot-dashed line), constant star formation (solid line), and exponentially increasing star formation (dashed line), and three metallicities: $Z=10^{-4},\ 10^{-3}, 10^{-2}$. The underlying spectral energy distributions assume no escape of Lyman-continuum photons (and hence include nebular emission) but have no dust attenuation/reddening.\label{fig:theory_sfh}}
\end{figure}

We begin by exploring the impact of the star formation history. In the top panel of Figure \ref{fig:theory_sfh} we show the break strength as a function of the \emph{duration} of star formation assuming  four different star formation histories (SFH): an instantaneous burst, an exponentially declining ($\tau=-100$ Myr) SFH, a constant SFH, and an exponentially increasing ($\tau=100$ Myr) SFH. In each case we assume no escape of LyC photons (i.e. maximal nebular emission) and no dust attenuation. In the bottom panel we instead express the $x$-axis in terms of the median (mass-weighted) age of the composite stellar population. The break strength broadly increases with increasing median age and SF duration. However, the relationship between age and break strength is sensitive to the shape of the recent star formation history. Models with recent star formation (i.e. constant or exponentially increasing) typically have bluer breaks for the same median age or duration. This immediately raises concerns about using the break strength as a diagnostic of the age for star forming galaxies, at least in isolation. 

\subsection{Metallicity}\label{sec:theory:Z}

In Figure \ref{fig:theory_sfh} we also show predictions for three different metallicities: $Z=0.0001$, $0.001$, and $0.01$. The break has a complex but relatively modest ($<0.1$ dex) metallicity dependence, which itself is also age dependent.
This behaviour is driven, at least in part, by the contribution of nebular continuum emission. 
For young ($<10$ Myr) stellar populations the break is stronger for lower-metallicity populations but at older ages our intermediate metallicity ($Z=0.001$) scenario gives rise to slightly stronger break.

\subsection{Dust}\label{sec:theory:dust}

For a monotonically declining attenuation curve the break and UV continuum slope will be sensitive to dust. As dust attenuation is increased the break strength will increase (redden). This is sensitive to the shape of the attenuation curve, but $\tau_{V}=1$ dust attenuation assuming a simple $\tau\propto\lambda^{-1}$ curve yields an increase in the break strength of $\approx 0.1$, comparable to significantly changing the metallicity or a 0.5 dex increase in age (for a instantaneous star formation history). Since dust may preferentially affect galaxies with ongoing star formation the impact would be to narrow the distribution of break strengths.

\subsection{Lyman continuum escape fraction}\label{sec:theory:fesc}

\begin{figure}
	\includegraphics[width=\columnwidth]{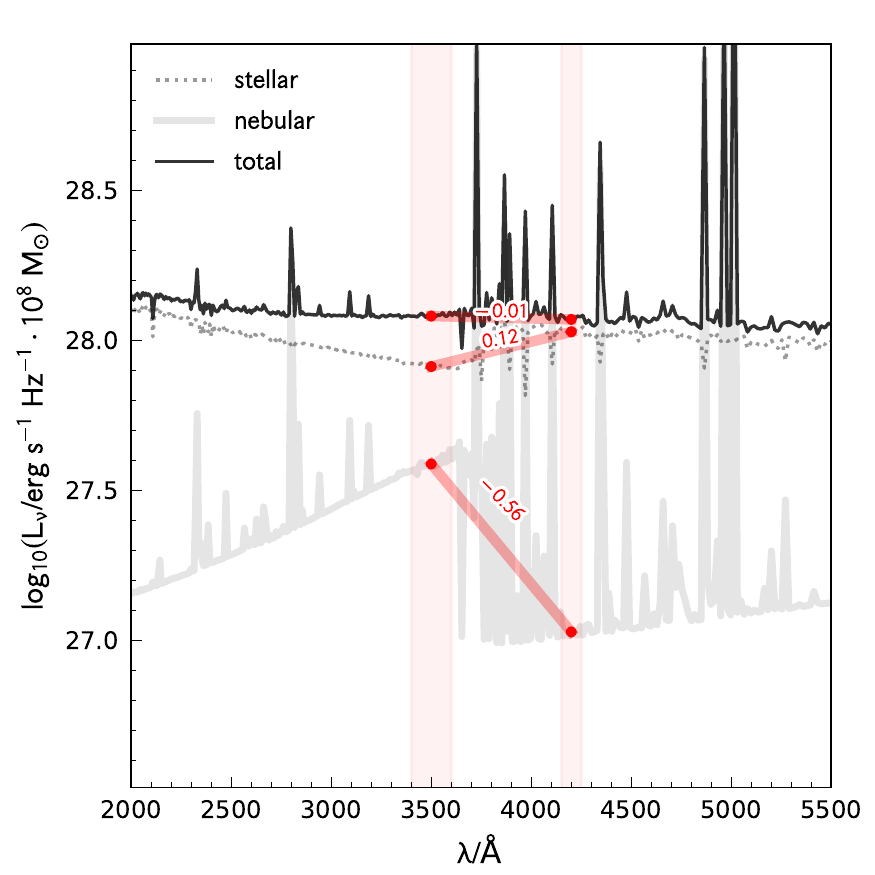}
	\caption{Stellar, nebular, and total spectral energy distributions for a dust-free star forming galaxy (scenario B in Figure \ref{fig:theory_sed}) along with the strength of the Balmer break. \label{fig:sed_neb}}
\end{figure}

\begin{figure}
	\includegraphics[width=\columnwidth]{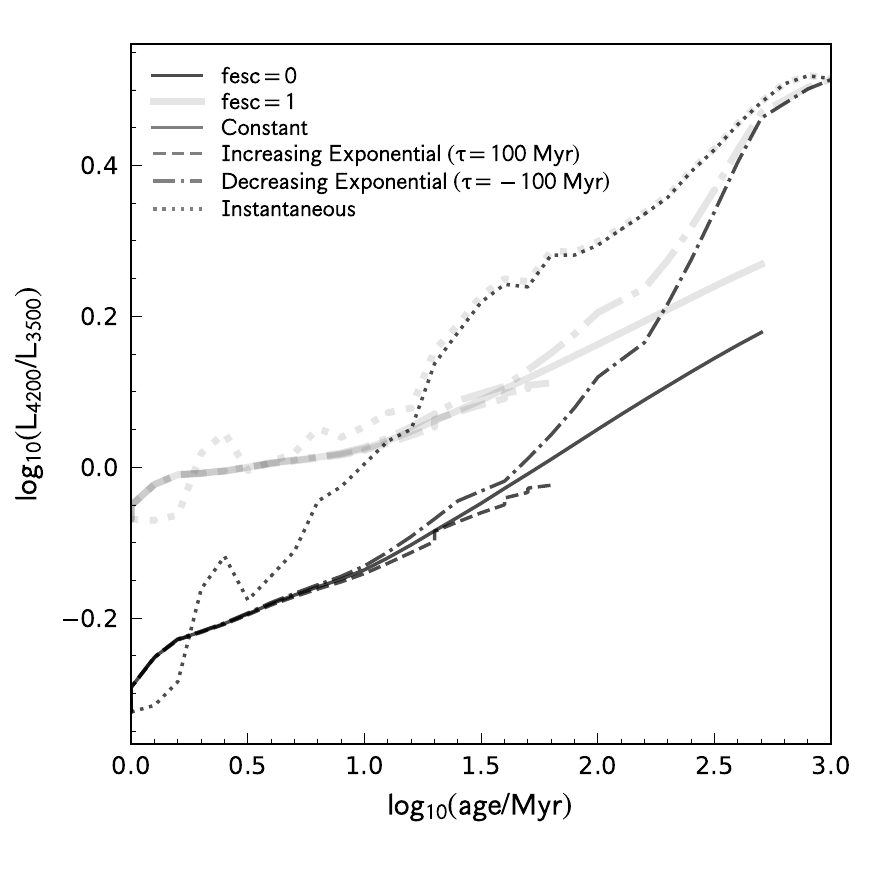}
	\caption{The sensitivity of the break strength to the star formation history and Lyman continuum escape fraction assuming $Z=0.005$.\label{fig:theory_fesc}}
\end{figure}

While our definition of the break is chosen to minimise the impact of nebular line emission it is still susceptible to nebular continuum emission. Figure \ref{fig:sed_neb} shows the pure stellar, nebular, and total (composite) spectrum assuming a 100 Myr constant star formation history and no dust. While the intrinsic (stellar) spectrum implies a break strength of $\approx 0.1$, the addition of nebular continuum emission, which has a sharp negative ($\approx -0.56$) break, results in a shallower total break $\approx 0$. Since the contribution of nebular emission is sensitive to the star formation history, and in particular the age, the impact of nebular emission is similarly sensitive. Figure \ref{fig:theory_fesc} shows the break strength assuming both $f_{\rm esc}=0$ and $f_{\rm esc}=1$. Nebular emission decreases the break strength by up-to 0.2 dex for young ages, with the magnitude dropping for longer durations of star formation. Without a robust estimate of the contribution of nebular emission this suggests it would be extremely challenging to constrain accurate ages for galaxies with ages less than a few tens of Myr.

\subsection{Choice of Initial Mass Function}\label{sec:theory:imf}

\begin{figure}
	\includegraphics[width=\columnwidth]{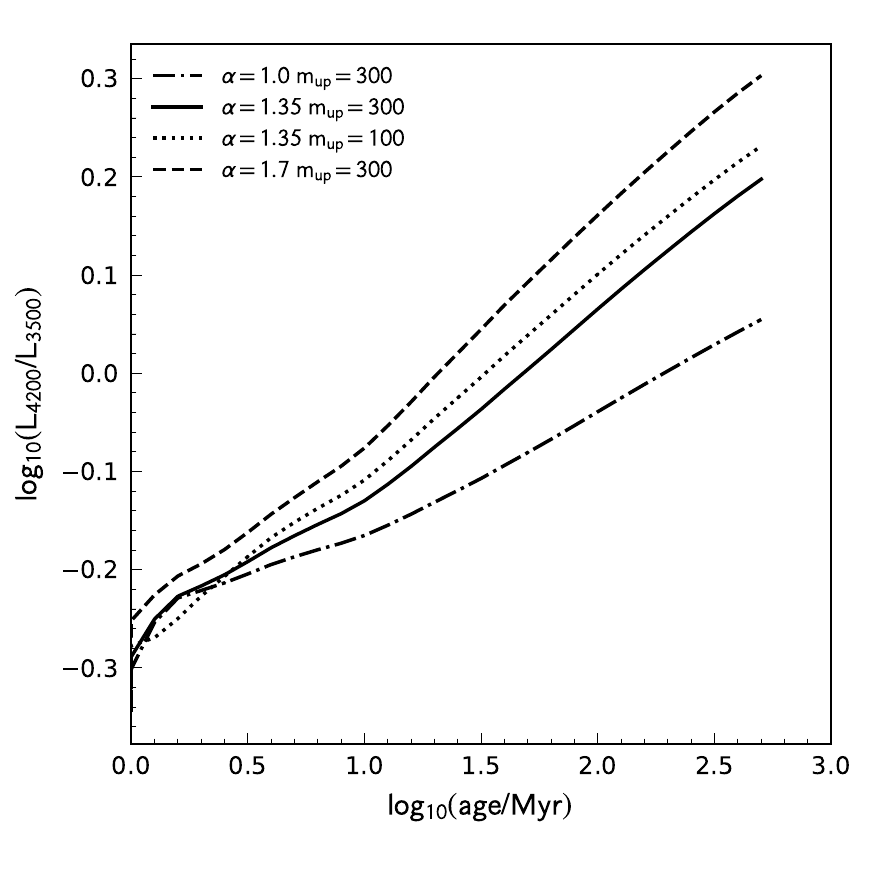}
	\caption{The sensitivity of the break strength to the age and choice of initial mass function (IMF) assuming a constant star formation history, $f_{\rm esc}=0$, $Z=0.005$, and the BPASS SPS model. Four IMFs are shown: three with different values of the high-mass slope $\alpha_3\in\{2, 2.35, 2.7\}$ and one with a $\alpha_3=2.35$ but lower upper-mass cut-off. \label{fig:theory_imf}}
\end{figure}

Any model spectral energy distribution is also sensitive to the choice of initial mass function (IMF). This is particularly relevant in the high-redshift Universe where the IMF is very uncertain \citep[e.g][]{Steinhardt22,Bate23}. While \flares\ assumes a constant IMF it is useful to explore the impact of the IMF in the context of our simple toy model. Figure \ref{fig:theory_imf} shows the impact of changing the IMF to four alternatives: an IMF with a shallower ($\alpha_{3}=1$) high-mass slope than Salpeter ($\alpha_3=1.35$), a Salpeter slope\footnote{In our previous and following analysis we assume the \citet{chabrier_galactic_2003} IMF. This has a slightly shallower high-mass slope slope than Salpeter.}, a Salpeter slope but with a lower high-mass cut ($m_{\rm up}=100\ {\rm M_{\odot}}$), and a steeper slope ($\alpha_3=1.7$). In the first case the result is more high-mass stars present producing a bluer break. For 100 Myr constant star formation duration the slope is $\approx 0.1$ dex smaller. Conversely assuming the steeper IMF results in a redder break, again by $\approx 0.1$ dex assuming 100 Myr constant star formation. Reducing the high-mass cut off has a smaller effect, increasing the break strength by $\approx 0.025$.

\subsection{Choice of Stellar Population Synthesis Model}\label{sec:theory:sps}

\begin{figure}
	\includegraphics[width=\columnwidth]{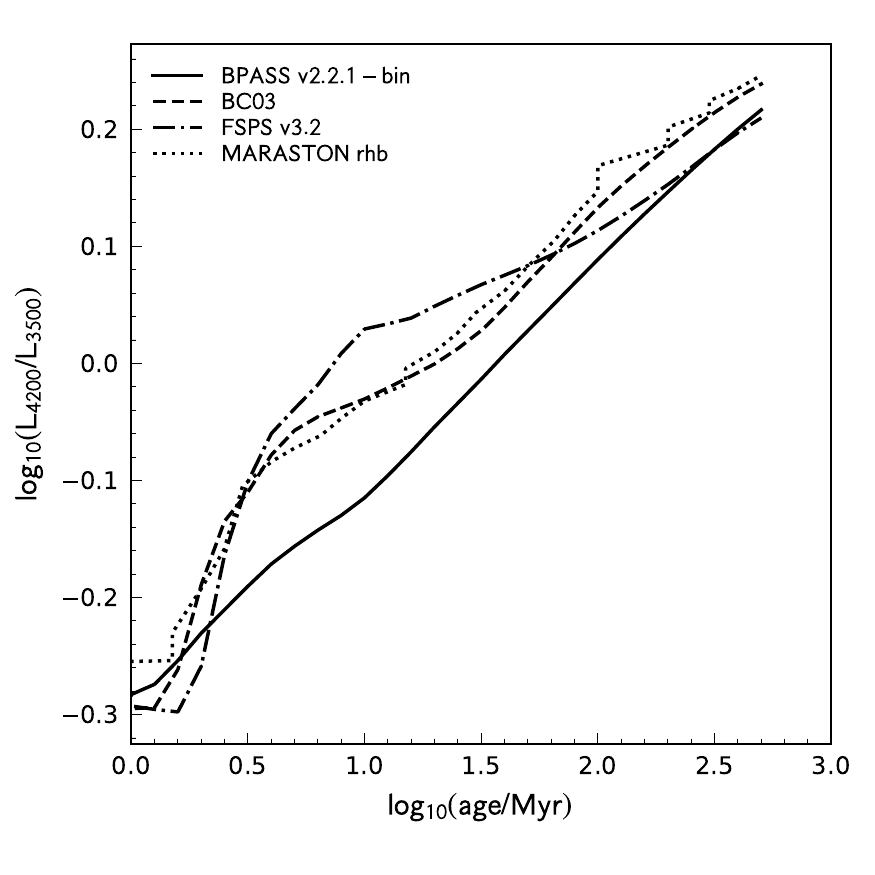}
	\caption{The sensitivity of the break strength to the duration of star formation and choice of stellar population synthesis model assuming a constant star formation history, $f_{\rm esc}=0$ and $Z=0.005$. Four models are considered: BPASS v2.2.1 (our default), \citet{chabrier_galactic_2003}, the Flexible Stellar Population Synthesis v3.2 model \citep{fsps}, and the \citet{Maraston} model. \label{fig:theory_sps}}
\end{figure}

The predicted spectral energy distributions of stellar populations are also sensitive to the choice of assumed stellar population synthesis (SPS) model.  Since different SPS models follow stellar evolution and atmospheres in distinct ways, discrepancies between their predictions can occasionally occur \citep[e.g][]{Byrne23}. To gauge the impact of this choice on our predictions in Figure \ref{fig:theory_sps} we show how the choice of stellar population synthesis model affects the break strength for three different models in addition to our default BPASS v2.2.1: \citet{BC03}, the \texttt{Flexible Stellar Population Synthesis} \citep[FSPS,][]{fsps} v3.2 model, and the \citet{Maraston} model. For ages 100$-$1000 Myr these models agree within 0.05 dex, however for younger ages ($\sim 10$ Myr) we find that BPASS yields significantly bluer ($\approx -0.1$) breaks.

%% file: Sections/flares.tex
\section{Predictions from the First Light And Reionisation Epoch Simulations}\label{sec:flares}

In this study we make use the First Light And Reionisation Epoch Simulations (\flares), introduced in  \citet{FLARES-I} and \citet{FLARES-II}, and we defer the reader to that paper for full details. In short, \flares\ is a suite of hydrodynamical re-simulations utilising the AGNdT9 variant of the \eagle\ simulation project \cite[][]{schaye2015_eagle,crain2015_eagle} with identical resolution as the reference simulations. The core \flares\ suite consists of 40 $14/h\ {\rm cMpc}$ radius re-simulations of regions selected from a large $(3.2\ {\rm cGpc})^3$ dark matter only simulation \citep{CEAGLE}. The selected regions span a large range in over-density (at $z\approx 4.7$): $\delta + 1 \approx -1\to 1$ with over-representation of the extremes of the density distribution. This strategy yields galaxies stretching over a larger range in environment, mass, and luminosity than possible with a periodic volume using the same computational resources. The simulated sample of galaxies, which stretched from $M_{\star}=10^8-10^{11}\ {\rm M_{\odot}}$ (at $z=5$), is well aligned with all but the most sensitive observations possible with \jwst.

\subsection{Spectral Energy Distribution modelling}\label{sec:methods::sed}

The spectral energy distribution (SED) modelling of galaxies in \flares\ is described in depth in \citet{FLARES-II}. This builds on the work of \citet{Wilkins2016b} and \citet{Bluetides_dust} and again we defer the reader to these papers for the details of our methodology. In short, we associate every star particle in the simulation with a pure stellar SED based on its mass, age, and metallicity using v2.2.1 of the BPASS \citep{BPASS2.2.1} SPS library and assume a \citet[]{chabrier_galactic_2003} IMF. Star particles are then associated with an ionisation bound H\textsc{ii} region calculated using version 17.01 of the \textsc{cloudy} photo-ionisation code \citep{Cloudy17.02}. Specifically, we use the pure stellar spectrum as the incident radiation field, assuming a spherical geometry and that the total metallicity (but not detailed abundances) of the nebula is identical to the star particle. We also assume a covering fraction of 1 (corresponding to a LyC escape fraction of $\approx 0$ for an ionisation bound nebula), a hydrogen density of $10^{2}\ \rm{cm}^{-3}$, and a metallicity and age dependent volume-averaged ionisation parameter referenced at $t=1$ Myr and $Z=0.01$ of $<U>=0.01$. Since this work is focused on a continuum feature specifically chosen to avoid strong line emission, our results are not strongly sensitive to these choices other than LyC escape fraction. Following this, we associate young stellar populations \citep[with age less than 10 Myr, following][that birth clouds disperse along these timescales]{CF00} with a dusty birth cloud. For every star particle we then determine the line-of-sight surface density of metals $\Sigma_Z$ in the $x,y$ plane and use this to determine an inter-stellar medium (ISM) dust attenuation.

%% file: Sections/predictions.tex
\subsection{Trend with Mass, Luminosity, and Redshift}\label{sec:predictions}

\begin{figure*}
    \includegraphics[width=2\columnwidth]{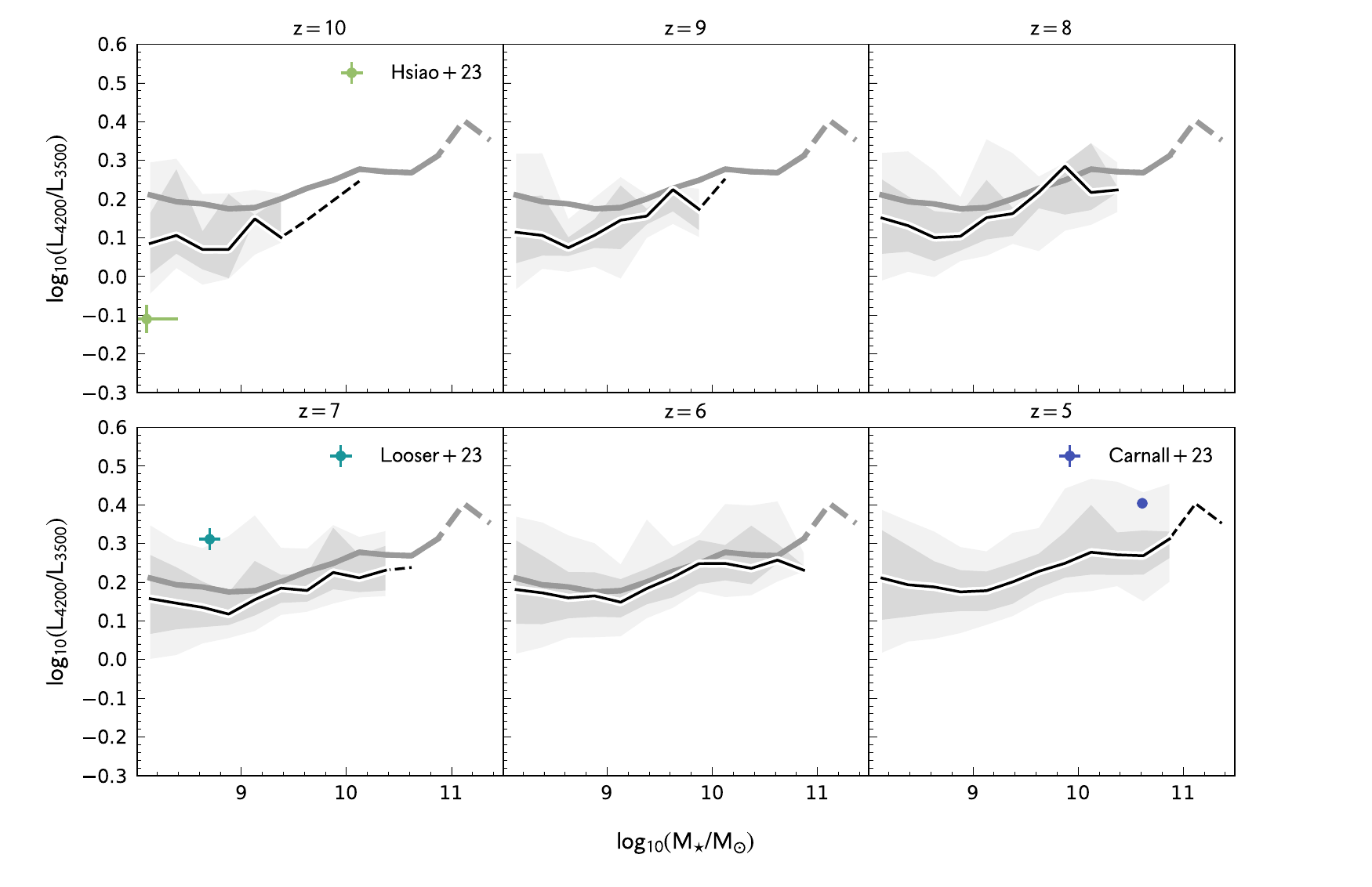}
    \includegraphics[width=2\columnwidth]{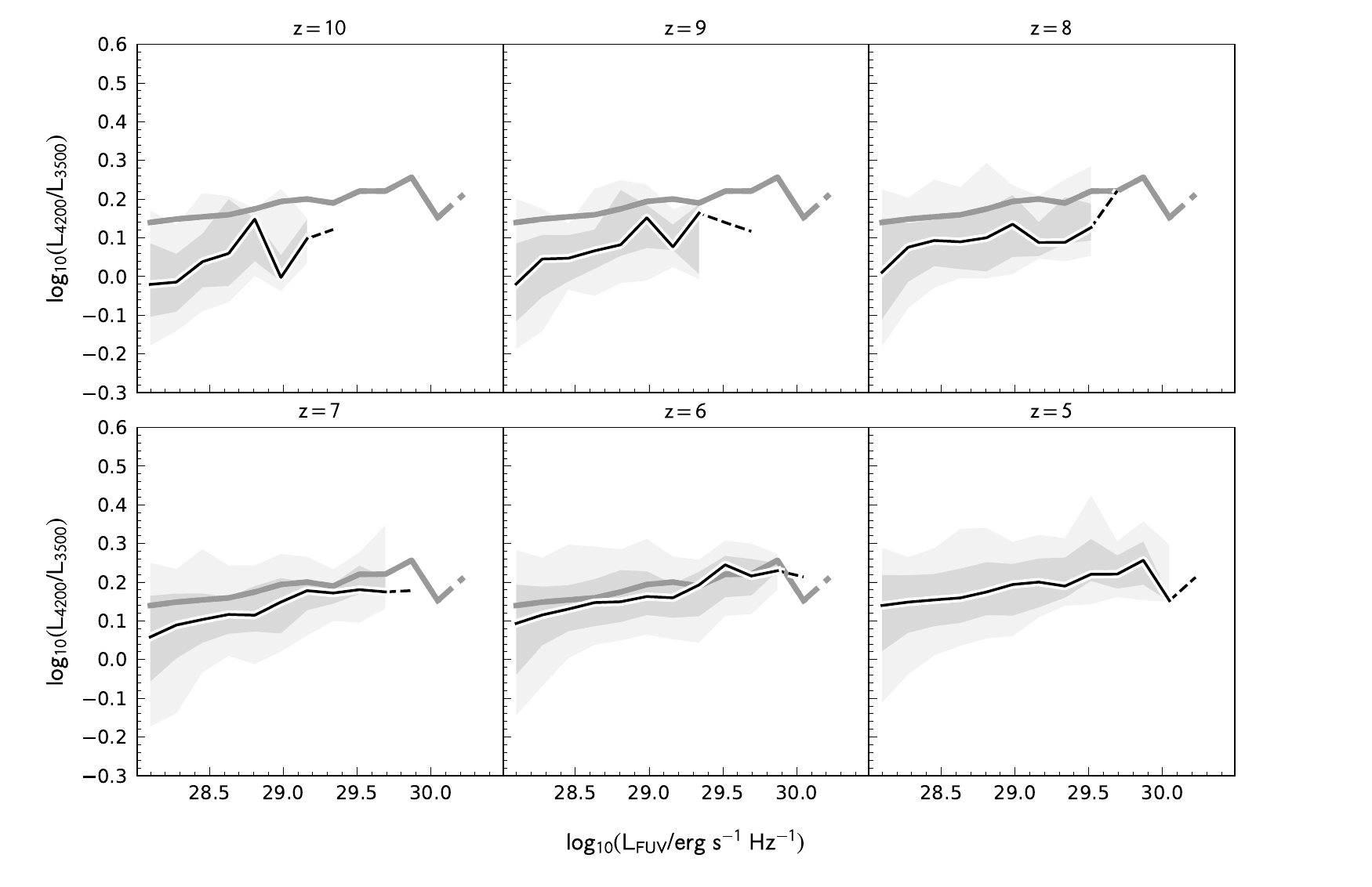}
	\caption{The predicted break strength as a function of the stellar mass (top) and attenuated rest-frame far-UV luminosity (bottom). The outlined dark line shows the median break strength while the two shaded regions show the 15.8-84.2th and 2.2-97.8th percentile ranges. The dashed line show the median when the number of galaxies in the bin falls below 10. The thick line shows the $z=5$ median relation to highlight the redshift evolution.  \label{fig:LM}}
\end{figure*}

We begin, in Figure \ref{fig:LM}, by presenting the Balmer break as a function of the rest-frame far-UV luminosity and stellar mass, respectively, for $z\in\{5,6,7,8,9,10\}$. Firstly, this reveals a clear redshift evolution with the break strength increasing by $\approx 0.1$ dex from $z=9$ to $z=5$. Secondly, the break strength also increases with luminosity, by $\approx 0.1$ dex from $L=10^{28}\to 10^{30}\ {\rm erg\ s^{-1}\ Hz^{-1}}$ at $z=5$.

\begin{figure}
    \includegraphics[width=\columnwidth]{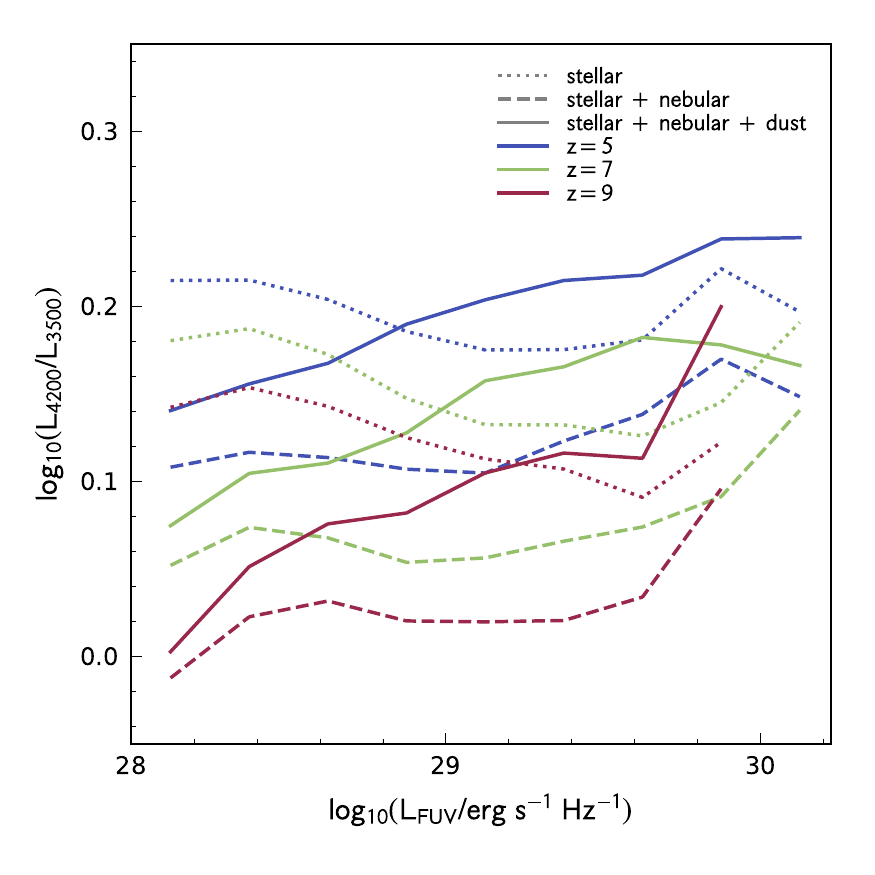}
	\caption{The impact of reprocessing by dust and gas on the median break strength at $z\in\{5,7,9\}$. The three sets of lines show predictions assuming no-reprocessing (pure stellar, dotted line), including nebular emission (dashed line), and including both dust and nebular emission (solid lines). \label{fig:reprocessing}}
\end{figure}

To gain further insight into the origin of these trends, in Figure \ref{fig:reprocessing} we present predictions at $z\in\{5,7,9\}$ both without dust attenuation alone (dashed line) and without both dust attenuation and nebular emission (i.e.~pure stellar breaks, dotted line) compared to our full predictions. The predicted pure stellar breaks evolve with redshift but remain relatively flat with UV luminosity. This agrees with \citet{FLARES-VII} who found that average ages only evolved subtly with stellar mass. While \citet{FLARES-VII} showed that there was strong evolution in metallicities the modelling in \S\ref{sec:theory:Z} demonstrated that for typical star formation histories in \flares\ the impact of metallicity is limited. 

We next explore the impact of nebular continuum emission. As noted in \S\ref{sec:theory:fesc}, nebular continuum emission is very blue across the break. Its inclusion in the \flares\ modelling shifts predicted average break colours by $-0.1$ dex, comparable to that found using our simple toy model in \S\ref{sec:theory:fesc}. Since the impact of nebular emission is sensitive to the star formation history, its impact is also redshift dependent, with galaxies at $z=9$ more strongly affected than those at $z=5$ due to the younger ages in the former. 

Finally, we explore the impact of dust attenuation. In \flares\ dust strongly correlates with intrinsic UV luminosity and stellar mass with a weaker correlation with attenuated UV luminosity \citep{FLARES-II, FLARES-III}. Since dust attenuation will redden/increase the break, this results in larger predicted breaks in the most luminous galaxies. 

In summary, pure stellar break strengths show no correlation with stellar mass and only mild redshift evolution ($\approx0.06$ dex, $z=9\to 5$). The inclusion of dust and nebular emission introduces a mass dependence, and strengthens the redshift evolution to $\approx0.1$ dex ($z=9\to 5$). 
 
Our results can be compared to those of \citet{Binggeli19}, who derived Balmer break statistics for $M_\star \geq 10^8 \ M_\odot$ galaxies using the \citet{Shimizu16}, FIRE-2 \citep{Ma19} and FirstLight \citep{Ceverino18} simulations, albeit for a more limited redshift range ($z=7$--9) than presented here. For galaxies in the stellar mass range $ 10^8$--$10^9\ M_\odot$  where \citet{Binggeli19} has sufficiently many simulated objects at all redshifts, the median Balmer break strength agrees with our predictions to within 0.05 dex at $z=7$, 8 and 9. However, the FIRE-2 simulations appear to produce a wider distribution of Balmer breaks at each redshift, with more outliers at both $\log_{10} (L_{4200}/L_{3500}) < 0$ and $> 0.3$.

\subsection{Correlation with physical properties}\label{sec:predictions:physical}

\begin{figure}
    \includegraphics[width=\columnwidth]{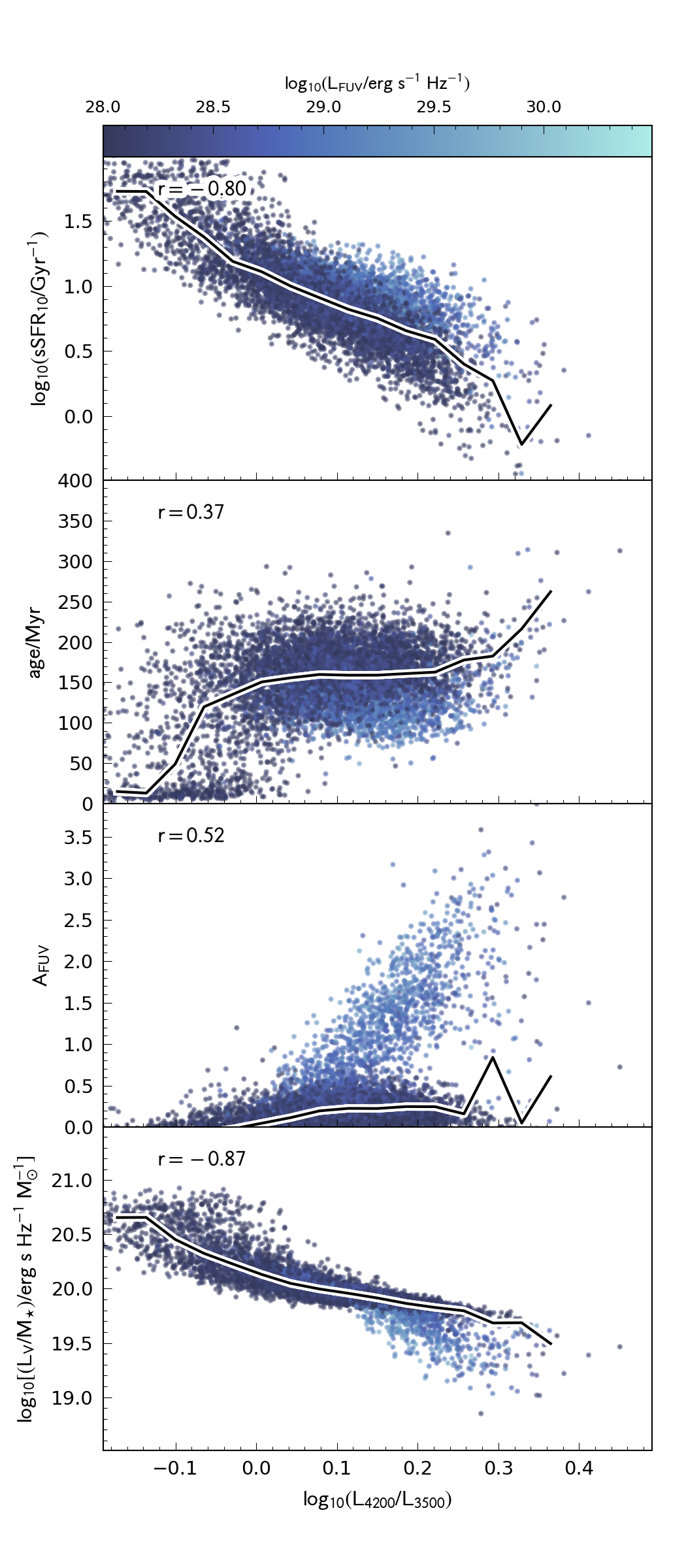}
	\caption{The correlation between the predicted break strength and the specific star formation rate, age, dust attenuation, and V-band mass-to-light ratio for galaxies at $z=5$. Individual objects are colour-coded by the rest-frame far-UV luminosity. The black line shows the median value. For each quantity and diagnostic the Pearson correlation coefficient $r$ is also presented. \label{fig:physical}}
\end{figure}

We now explore how the predicted break strength correlates with key physical properties, including the specific star formation rate, age, dust attenuation, and the rest-frame optical mass-to-light ratio. These relationships are shown in Figure \ref{fig:physical}. 

Firstly, we find a clear strong ($r=-0.8$) inverse relationship between the break strength and specific star formation rate. Here the star formation rate is measured on a timescale of 10 Myr but this assumption does not make a significant difference to the correlation. The correlation with the mass-weighted median age is significantly weaker ($r=0.37$). For the bulk of galaxies, which have break strengths $0.0-0.2$ there is no correlation with age. However, galaxies with the largest breaks tend to be older and vice versa. This lack of correlation reflects two issues: firstly, the break is also sensitive to the shape of the star formation history as well as other properties (nebular emission and dust) but also that galaxies in \flares\ exhibit a relatively tight range of ages \citep{FLARES-VII}. The correlation with dust attenuation is more complex. Most galaxies in \flares\ (with $L_{\rm FUV}>10^{28}\ {\rm erg\ s^{-1}\ Hz^{-1}}$) have relatively low attenuation \citep[i.e. $A_{\rm FUV}<0.5$, see][]{FLARES-II}. However, the most intrinsically luminous galaxies in \flares\ have increasing attenuation. Since increasing dust attenuation results in larger breaks there is a branch of increasingly dusty galaxies with increasing break strengths. Thus while the dustiest galaxies all have strong breaks, the break itself is not an unambiguous tracer of dust attenuation. 

Finally, we also explore the correlation of the break strength with the optical rest-frame $V$-band mass-to-light ratio. This reveals a strong positive ($r=0.87$) correlation. Galaxies with the strongest breaks thus tend to have larger stellar masses for the same $V$-band luminosity. This arises due to sensitivity of both the break and mass-to-light ratio to dust attenuation and the star formation history. This reinforces the utility of the break as a diagnostic of stellar masses.

\subsection{Comparison with observational direct constraints}\label{sec:predictions:observations}

At the time of writing, the first spectroscopic measurements of galaxies at $z>5$ made by \jwst\ are emerging, including the detection of the continuum at sufficiently high signal-to-noise to enable the measurement of the break \citep[e.g][]{Carnall2023, Bunker23, Cameron23, Looser23, Hsiao2023}. Here we compare our predictions with those studies that have published \citep{Hsiao2023} or made available \citep[via private communcation][]{Carnall2023, Looser23} observational constraints on the Balmer break.

\citet{Carnall2023} present NIRSpec observations of a massive quiescent galaxy (GS-9209) at $z\approx 4.7$ detecting the rest-frame 3000-9000\AA\ continuum at high signal-to-noise.This object has a measured break of $\log_{10}(L_{4200}/L_{3500})\approx 0.4$. This places it at the extreme end, but within the range of our predictions. Since GS-9209 was initially selected as an extremely red, and thus potentially quiescent source \citep{Caputi2004}, its location at the extreme end of our predicted distribution is unsurprising. 

\citet{Looser23} report the discovery of JADES-GS-z7-01-QU, a potentially quiescent galaxy at $z=7.3$ from JWST Advanced Deep Extragalactic Survey (JADES) observations. The break strength of this source $\approx 0.3$, though this value may be overestimated as the red window coincides with noisy features in spectrum. Nevertheless, like GS-9209, JADES-GS-z7-01-QU exhibits a strong break, lying at the extreme end of the predicted distribution. However, GS-9209 and JADES-GS-z7-01-QU are unlikely to be representative of the wider population of high-redshift galaxies and it therefore premature to make any inferences about the model.

\citet{Hsiao2023} present NIRSpec observations of MACS0647-JD, the triply-lensed source found to be at $z\approx 10.2$. Unlike GS-9209 and JADES-GS-z7-01-QU, MACS0647-JD is not selected to be quiescent and consistent with being a typical star forming galaxies. For MACS0647-JD \citet{Hsiao2023} report $M_{\star}=10^{8.1}\ {\rm M_{\odot}}$ and $\log_{10}(L_{4200}/L_{3500})\approx -0.11$. In contrast to GS-9209 and JADES-GS-z7-01-QU, MACS0647-JD lies \emph{below} our predictions and in fact just outside the central 95\% range. However, since this is a single object, and only marginally inconsistent, it is again impossible draw firm inferences about \flares.

These and other recent observations demonstrate \jwst's extraordinary potential to probe the rest-frame spectral energy distributions of galaxies in the distant Universe. With many more spectroscopic observations planned over the coming cycles \jwst\ will soon place strong constraints on the break providing a new tool to test galaxy formation models like \flares.

%% file: Sections/conclusion.tex
\section{Conclusions}\label{sec:conc}

In this work we have explored predictions for the strength of the Balmer/4000~\AA\ break predicted by the First Light And Reionisation Epoch Simulations (\flares) project. Prior to this we explored how various parameters (star formation and metal enrichment history, dust attenuation, and LyC escape fraction) and assumptions (stellar population synthesis model and initial mass function) affect the break using a simple toy model. 

\begin{itemize}

    \item Our toy modelling reveals that while the strength of the break is sensitive to the age (or duration) it is also sensitive to the shape of the star formation history, metallicity, dust attenuation, and LyC (ionising photon) escape fraction. The impact of both the shape of the SFH and LyC escape fraction can shift the break strength by $\sim 0.2$ dex. Metallicity is less important but can drive differences of up to $\approx 0.1$ dex.

    \item This modelling also reveals the sensitivity of the break strength to the choice of stellar population synthesis (SPS) model and initial mass function (IMF). The choice of the SPS model can change the break strength by $>0.1$ dex, with the largest impact at star formation durations of $\sim 10$ Myr. The impact of the IMF grows with increasing duration of star formation, increasing to up-to 0.3 dex for $\alpha_3 = 1.0\to 1.7$ after a few hundred Myr of star formation.

    \item \flares\ predicts a flat relationship between attenuated far-UV luminosity or mass and the \emph{intrinsic} break strength. Including nebular emission reduces the break strength by $\approx 0.1$ dex while including dust increases the break strength and produces a slightly positive relationship between attenuated far-UV luminosity or mass and the break strength. The average break strength also varies with redshift, increasing by around 0.1 dex from $z=9\to 5$.

    \item The break strength inversely correlates with both specific star formation rate and optical mass-to-light ratio but only weakly correlates with the average age, and then only at the extremes. 

    \item At the time of writing there have only been a handful of robust spectroscopic measurements of the break strength and these are biased samples. However, this will soon change, ultimately providing a new constraint on galaxy formation models in the distant Universe.
    
\end{itemize}